\documentclass[journal]{IEEEtran}
\IEEEoverridecommandlockouts

\usepackage[tbtags]{amsmath}
\usepackage{amssymb}
\usepackage{algpseudocode}
\usepackage{algorithm}
\usepackage{enumitem}
\usepackage{array}
\usepackage{bm}
\usepackage{tikz}
\usepackage{color,soul}
\usetikzlibrary{shapes,shapes.geometric,arrows,fit,calc,positioning,automata}
\usepackage[caption=false]{subfig}
\usepackage[hidelinks]{hyperref}

\newcommand\srm[2]{#1_\mathrm{#2}}

\newcommand{\um}{\bm{u_\mathrm{m}}}
\newcommand{\ub}{\bm{u_\mathrm{b}}}
\newcommand{\xm}{\bm{x_\mathrm{m}}}
\newcommand{\xb}{\bm{x_\mathrm{b}}}
\newcommand{\xmb}{\bm{x_\mathrm{m/b}}}
\newcommand{\ym}{y_\mathrm{m}}
\newcommand{\yb}{y_\mathrm{b}}
\newcommand{\ymb}{y_\mathrm{m/b}}
\newcommand{\umb}{\bm{u_\mathrm{m/b}}}
\newcommand{\Umb}{\bm{\mathcal U_\mathrm{m/b}}}
\newcommand{\creflect}{c_\mathrm{ref}}
\newcommand{\cexpand}{c_\mathrm{exp}}
\newcommand{\ccontract}{c_\mathrm{con}}
\newcommand{\cshrink}{c_\mathrm{shrink}}
\newcommand{\xc}{\bm{\srm{x}{c}}}
\newcommand{\Vmin}{\srm{V}{min}}

\newcommand{\jmax}{j_\mathrm{max}}

\newcommand{\cEma}{\alpha}

\newcommand{\E}{\mathrm{E}}
\newcommand{\fmin}{\mu_\mathrm{min}}

\newcommand{\fs}{f_\mathrm{s}}

\newcommand{\sn}{\sigma_\mathrm{n}}

\newcommand{\tf}{t_\mathrm{f}}
\newcommand{\xbest}{\bm{x_\mathrm{best}}}
\newcommand{\xlb}{\bm{x_\mathrm{lb}}}
\newcommand{\xub}{\bm{x_\mathrm{ub}}}
\newcommand{\xlbabs}{\bm{x_\mathrm{lb, abs}}}
\newcommand{\xubabs}{\bm{x_\mathrm{ub, abs}}}

\newcommand{\zmin}{z_\mathrm{min}}
\newcommand{\zmax}{z_\mathrm{max}}

\newcommand{\Fsp}{F_\mathrm{e}}
\newcommand{\Ff}{F_\mathrm{f}}
\newcommand{\Fmag}{F_\mathrm{mag}}
\newcommand{\ksp}{k_\mathrm{sp}}
\newcommand{\zsp}{z_\mathrm{sp}}
\newcommand{\cf}{c_\mathrm{f}}
\newcommand{\bksp}{ \hat k_\mathrm{sp}}
\newcommand{\bcf}{ \hat c_\mathrm{f}}

\newcommand{\Relg}{\mathcal R_\mathrm{g}}

\newcommand{\Relc}{\mathcal R_\mathrm{c}}
\newcommand{\dRelg}{\mathcal R'_\mathrm{g}}
\newcommand{\ddRelg}{\mathcal R''_\mathrm{g}}

\newcommand{\mmov}{m_\mathrm{mov}}
\newcommand{\ucoil}{\upsilon_\mathrm{coil}}
\newcommand{\icoil}{i_\mathrm{coil}}
\newcommand{\ieddy}{i_\mathrm{eddy}}
\newcommand{\keddy}{k_\mathrm{eddy}}

\newcommand{\zf}{z_\mathrm{f}}

\newcommand{\thetanomi}{\theta_i^\mathrm{nom}}

\newcommand{\xtol}{\mathcal U_\mathrm{tol}}

\newcommand{\sigtheta}{\sigma_p}

\newcommand{\pnomi}{p^\mathrm{nom}_i}

\newcommand \lfrac[2]	{ #1 / #2 }

\DeclareMathOperator*{\argmax}{arg\,max}
\DeclareMathOperator*{\argmin}{arg\,min}

\DeclareMathOperator*{\diag}{diag}

\algnewcommand{\LeftComment}[1]{\; \(\triangleright\) #1}
\makeatletter
\algnewcommand{\LineComment}[1]{\Statex \hskip\ALG@thistlm \(\triangleright\) #1}
\makeatother
\algnewcommand\algorithmicswitch{\textbf{switch}}
\algnewcommand\algorithmiccase{\textbf{case}}
\algdef{SE}[SWITCH]{Switch}{EndSwitch}[1]{\algorithmicswitch\ #1}{\algorithmicend\ \algorithmicswitch}
\algdef{SE}[CASE]{Case}{EndCase}[1]{\algorithmiccase\ #1}{\algorithmicend\ \algorithmiccase}
\algtext*{EndSwitch}
\algtext*{EndCase}

\title{\LARGE \bf
	Run-to-run control with Bayesian optimization for soft landing of short-stroke reluctance actuators
}

\author{Eduardo Moya-Lasheras and Carlos Sagues
	\thanks{This work was supported in part by the Ministerio de Ciencia Innovaci\'on y Universidades, Gobierno de Espa\~na - European Union under project RTC-2017-5965-6 of subprogram Retos-Colaboraci\'on, in part by the Gobierno de Arag\'on-FSE 2014-20, and in part by project DGA-T45\_17R/FSE. (Corresponding author: Eduardo Moya-Lasheras.)}
	\thanks{E. Moya-Lasheras and C. Sagues are with the Departamento de Informatica e Ingenieria de Sistemas (DIIS) and the Instituto de Investigacion en Ingenieria de Aragon (I3A), Universidad de Zaragoza, Zaragoza 50018, Spain (email: emoya@unizar.es, csagues@unizar.es).}
  \thanks{\textcolor{red}{This is the accepted version of the manuscript: E. Moya-Lasheras and C. Sagues, ``Run-to-Run Control With Bayesian Optimization for Soft Landing of Short-Stroke Reluctance Actuators," in IEEE/ASME Transactions on Mechatronics, vol. 25, no. 6, pp. 2645-2656, Dec. 2020, doi: 10.1109/TMECH.2020.2987942. 
\textbf{Please cite the publisher's version}. For the publisher's version and full citation details see:\\
\protect\url{https://doi.org/10.1109/TMECH.2020.2987942}. 
}}

 \thanks{© 2020 IEEE.  Personal use of this material is permitted.  Permission from IEEE must be obtained for all other uses, in any current or future media, including reprinting/republishing this material for advertising or promotional purposes, creating new collective works, for resale or redistribution to servers or lists, or reuse of any copyrighted component of this work in other works.}
}

\begin{document}

\maketitle
\thispagestyle{plain}
\pagestyle{plain}

\begin{abstract}
	There is great interest in minimizing the impact forces of reluctance actuators during commutations, in order to reduce contact bouncing, acoustic noise and mechanical wear. In this regard, a run-to-run control algorithm is proposed to decrease the contact velocity, by exploiting the repetitive operations of these devices. The complete control is presented, with special focus on the optimization method and the input definition. The search method is based on Bayesian optimization, and several additions are introduced for its application in run-to-run control, e.g. the removal of stored points and the definition of a new acquisition function. Additionally, methods for the input parametrization and dimension reduction are presented. For analysis, Monte Carlo simulations are performed using a dynamic model of a commercial solenoid valve, comparing the proposed search method with two alternatives. Furthermore, the control strategy is validated through experimental testing, using several devices from the same ensemble of solenoid valves.
\end{abstract}

\section{Introduction}\label{introduction}
Reluctance actuators are electromechanical devices that rely on reluctance forces to change the position of their movable parts. Simple single-coil short-stroke reluctance actuators, e.g. electromagnetic relays or solenoid valves, are extensively used in on-off switching operations of electrical, hydraulic or pneumatic circuits. However, the range of applications is restricted because of one major drawback: the strong impact at the end of each commutation, which provokes contact bouncing, mechanical wear and acoustic noise. Mitigating those impacts is of great interest, as it potentially extends their service life, makes them operate more quietly, and opens them to applications with more stringent requirements. To decrease contact velocities of the moving parts in reluctance actuators, soft-landing control strategies must be designed.

One of the most straightforward approaches is to design feedback control strategies to track predefined position trajectories \cite{Eyabi2006,Mercorelli2012}. However, in most low-cost short-stroke reluctance actuators, there is no affordable and feasible method to measure the position in real time. To circumvent this limitation, estimators can be designed for the armature position \cite{Rahman1996}, or other position-dependent variables \cite{Ramirez-Laboreo2019a}. However, these solutions require precise models for each device that accounts for---among other aspects---its nonlinearities, discrete behavior, time-varying parameters, or measurement errors. Moreover, in many cases, there is a significant unit-to-unit variability among devices from the same ensemble, and the identification of each unit may impose a prohibitive cost.

Run-to-run (R2R) control, in contrast with other learning-type strategies, only requires one evaluation value for each cycle \cite{Wang2009}. It is ideal for controlling low-cost reluctance actuators because it does not require the position during operations. Instead, other variables are derived to evaluate each cycle, e.g. the bouncing duration \cite{Ramirez-Laboreo2017a} or the sound intensity of the impact.

Measurement-based batch optimization implementations can be explicit or implicit, depending on whether it requires a process model or not \cite{Srinivasan2002}. Designing an explicit R2R strategy for reluctance actuators is very challenging. Even if the system dynamics is modeled, the cost function---which maps the decision vector (input) with the evaluation variable (output)---is unknown. Without an explicit definition of the cost function, each evaluation requires querying at a certain decision point and actually commutating the device. As stated by \cite{Srinivasan2002}, the gradient can be approximated by disturbing the $d$ decision variables. However, the estimated derivatives are very sensitive to noise and need at least $d+1$ function evaluations in each iteration. A better solution for this type of problem, proposed by \cite{Ramirez-Laboreo2017a}, is based on a derivative-free pattern search method \cite{Rios2013}, but it still requires many function evaluations in each iteration.

Bayesian optimization is a well-known method of black-box global optimization. In each iteration, it approximates the black-box function with a random process regressor---typically Gaussian \cite{Rasmussen2006}---depending on data from previous iterations and, through the maximization of a utility function, selects the following points to be evaluated. It has been proven to be effective in real-time control applications, e.g. maximum power point tracking \cite{Abdelrahman2016}, or altitude optimization of airborne wind energy systems \cite{Baheri2017}.

Besides the optimization algorithm, the input signal selection is critical for the control performance. Furthermore, the signal must be parametrized so it can be modified in each iteration from a limited set of decision variables. While bang-off-bang voltage signals are adequate for soft landing and very easily parametrizable, the dynamics of the current is relatively slow and must be taken into account. Although the current parametrization is more challenging, this paper argues that it is a better choice for input than the voltage, because it is more robust to changes in the temperature.

This paper presents a R2R strategy with Bayesian optimization for soft-landing control of reluctance actuators. Several elements are introduced to the optimization algorithm: the limitation of stored previous data by means of the combination or removal of observations, the adaptability of the search bounds, and the definition of a new acquisition function. Another contribution is the input parametrization, which relates the decision variables to parameters of the dynamic model. To demonstrate the effectiveness of the proposal, multiple simulations are performed with a dynamic model that fits a commercial solenoid valve. The proposed R2R control is compared with two alternatives. The first one uses a pattern search method, and was previously proposed for bounce reduction of relays and contactors. The second one uses a Nelder-Mead search method, which requires fewer function evaluations in each iteration than the pattern search method. Then, to validate the control applicability in real devices, the R2R strategies are applied to solenoid valves using an experimental setup.

The preliminary ideas of the proposal were presented at the European Control Conference \cite{Moya-Lasheras2019a}. The most substantial changes are the addition of adaptive search bounds (Section \ref{sec: limit size}), the definition of a new acquisition function (Section \ref{sec: acquisition}), the model-based input parametrization (Section \ref{sec: input}), and the new simulations and experiments to support the contributions (Section \ref{sec: results}).

\section{Run-to-run control with Bayesian optimization}\label{sec: R2R}
\subsection{Main algorithm}
The generalized R2R control algorithm is presented. It must be iterative to account for and exploit the cyclic operations of the actuators. These devices are characterized by having two distinct operation types depending on the motion direction: making and breaking. These two types of operation act alternatively, so a complete commutation cycle consists of one operation of each. The R2R solution (see Algorithm \ref{alg: main}) is a loop in which every iteration $k$ comprises three steps.

The first step is the generation of the input signals for the making and breaking operations from a set of $d$ decision variables (\textsc{Generate Input}). Formally, it is expressed as
\begin{equation}\label{eq: U}
	\umb^k(t) = \Umb\big( \xmb^k, t \big),
\end{equation}
where $t$ denotes the time dependence, $\mathrm{m/b}$ serves as a distinction between making ($\mathrm{m}$) and breaking ($\mathrm{b}$) operations, $\umb^k(t)$ is the input signal, $\xmb^k \in \mathbb R^d$ is the vector of decision variables, and $\Umb$ is the input function. In order to define $\Umb$, the input must be parametrized.

The second step is the application of these signals and the observation of the costs $\ymb^k$. In general, they depend on their corresponding input, which in turn depends on $\xmb^k$,
\begin{equation}
	\ymb^k = \psi \left( \xmb^k \right) + \delta^k,
\end{equation}
where $\psi$ is the black-box cost function, and $\delta^k$ denotes the additive effect of system disturbances and measurement errors.

Then, the last step is the optimization process (\textsc{Search}), where the objective is to minimize the costs. The next decision vectors ($\xm^{k+1}$ and $\xb^{k+1}$) are obtained from previous data,
\begin{align}
	\bm{\mathcal{X}_\mathrm{m/b}}^{k} & =\left\{ \xmb^{i} \mid i = 1,\,2,\,\ldots,\,k \right\}, \\
	\mathcal{Y}_\mathrm{m/b}^{k}      & =\left\{ \ymb^{i} \mid i = 1,\,2,\,\ldots,\,k \right\}.
\end{align}

\begin{algorithm}[t]
	\caption{Run-to-run control}\label{alg: main}
	\begin{algorithmic}[1]
		\State \textbf{Initialize:} $\xm^1, \, \xb^1 $
		\For{$ k \leftarrow 1 $ \textbf{to} num. commutations}
		\State $\um^k(t) \leftarrow $ \Call{Generate Input}{$\xm^k$}
		\State $\ub^k(t) \leftarrow $ \Call{Generate Input}{$\xb^k$}
		\State Apply $\um^k(t)$ and measure $\ym^k$
		\State Apply $\ub^k(t)$ and measure $\yb^k$
		\State $\xm^{k+1} \leftarrow $\Call{Search}{$ \bm{\mathcal{X}_\mathrm{m}}^k, \, \mathcal{Y}^k_\mathrm{m} $} \vspace{0.25mm}
		\State $\xb^{k+1} \leftarrow $\Call{Search}{$ \bm{\mathcal{X}_\mathrm{b}}^k, \, \mathcal{Y}^k_\mathrm{b} $}
		\EndFor
	\end{algorithmic}
\end{algorithm}

Notice that the frequency of the cycles is limited by the computation time of the functions \textsc{Generate Input} and \textsc{Search}. If that time is not small enough, it is necessary to adapt the algorithm to work around this issue, e.g. by commutating the device several times in each iteration without updating the decision vectors, or by computing in parallel the function algorithms for the making and breaking operations.

While the function \textsc{Generate Input} depends on the choice of input signal and the type of actuator, the following description of optimization function SEARCH is generalized.

\subsection{Optimization method}\label{sec: bay opt 2}
The proposed \textsc{Search} function is based on Bayesian optimization, which is a method for finding the global optimum of an unknown function. From previous data, it builds a stochastic regressor of the black-box function and predicts the output for any point $\bm x$. The selection of the next point to be evaluated $\bm x^{k+1}$ is carried out by maximizing an acquisition function $\srm{f}{acqn}$ of the predicted output. The evaluation $y^{k+1}$ is obtained in the next iteration of the R2R Algorithm \ref{alg: main}.

The proposed function is described in Algorithm \ref{alg: BO main}. Its inputs are the current point (decision vector $\bm x^k$), which has been obtained in the previous iteration, and its evaluation $y^k$. Its output is the next point $\bm x^{k+1}$. Some parameters, e.g. the observation noise variance $\sn^2$, are set as constant. Also, there are some persistent variables, e.g. the number of stored points $j$, which are changed inside the function but are not required outside of it. Note that these variables are different for each operation type but, for the sake of simplicity, that distinction is omitted. The algorithm can be divided into three steps:

\begin{algorithm}[t]
	\caption{Optimization}\label{alg: BO main}
	\begin{algorithmic}[1]
		\Function{Search}{$ \bm x^k, \, y^k $}
		\State \textbf{Constant:} $ \sn^2, \, d, \, \jmax $
		\State \textbf{Persistent:} $\bm X, \, \bm Y, \, \bm \Sigma, \, j $
		\LineComment{Learning}
		\State $ j \leftarrow j+1 $
		\State $ \left( \bm X_{j}, \, Y_j, \, \bm \Sigma_{j,j} \right) \leftarrow \left( \bm x^k, \, y^k, \, \sn^2 \right) $
		\LineComment{Data size constraining}
		\State $ \left(\bm X, \, \bm Y, \, \bm \Sigma, \, j \right) \leftarrow $ \Call{Merge}{$ \bm X, \, \bm Y, \, \bm \Sigma, \, j $}
		\State $ \left(\xlb, \, \xub\right) \leftarrow $ \Call{Bounds}{$ \bm X, \, \bm Y, \, \bm \Sigma$}
		\If{$ j > \jmax $}
		\State $ \left( \bm X, \bm Y, \bm \Sigma, j \right) \leftarrow $ \Call{Remove}{$ \bm X, \bm Y, \bm \Sigma, j, \xlb, \xub$}
		\EndIf
		\LineComment{Acquisition}
		\State $ \bm x^{k+1} \leftarrow \argmax_{\xlb \leq \bm x \leq \xub} \srm{f}{acqn} \left( \bm x \, \left| \, \bm X, \bm Y, \bm \Sigma \right. \right) $
		\State \Return $ \bm x^{k+1} $
		\EndFunction
	\end{algorithmic}
\end{algorithm}

\begin{enumerate}[leftmargin=*]
	\item \textbf{Learning}. Updating the stored points $\bm X \in \mathbb R^{d \times j}$ and their evaluations $\bm Y\in \mathbb R ^{1 \times j}$ by the addition of the $k$th decision vector $\bm x^k$ and its cost $y^k$. The variance of the last observation $\sn^2$ is added to the covariance $\bm \Sigma$.
	\item \textbf{Data size constraining} ($j \leq \srm j {max}$). Observations are merged or removed if necessary. Furthermore, the search space is modified in each iteration. These processes are further discussed in Subsection \ref{sec: limit size}.
	\item \textbf{Acquisition}. Selection of next decision vector $\bm x^{k+1}$ by maximizing an acquisition function, given the previous data ($\bm X$, $\bm Y$ and $\bm \Sigma$). The search is restricted between the lower bound $\xlb$ and the upper bound $\xub$. The proposed acquisition function is defined in Subsection~\mbox{\ref{sec: acquisition}}.
\end{enumerate}

\subsection{Prior and posterior distributions}\label{subsec: dist}
The selected model for regression is the Gaussian process. It is the most popular one in the context of Bayesian optimization because it only requires simple algebraic operations to determine the corresponding posterior distribution. In general, it is completely specified by a mean $m(\bm x)$ and covariance function or kernel $k(\bm x, \bm x')$ \cite{Rasmussen2006},
\begin{equation}
	f(\bm x) \sim \mathcal{GP}\left(m(\bm x), k(\bm x, \bm x')\right).
\end{equation}

For convenience, $m$ is assumed to be constant. The chosen covariance function is squared exponential,
\begin{equation}\label{eq: kernel}
	\hspace{-0.3mm} k(\bm x, \bm x') = \sigma_f^2 \, \mathrm{exp}\left( -\frac{1}{2}\,(\bm x - \bm x')^\mathsf{T} \, \diag(\bm l)^{-2} \, (\bm x - \bm x') \right),
\end{equation}
where $\sigma_f^2$ is the characteristic variance and $\diag(\bm l) \in \mathbb{R}^{d \times d}$ is a diagonal matrix with the lengthscales for each dimension.

In a given iteration, we have the training outputs $\bm Y = f \! \left( \bm X \right) + \bm \varepsilon$. The output noise $\bm \varepsilon$ is an independently distributed Gaussian random vector whose covariance is the diagonal matrix $\bm \Sigma$. Given the properties of Gaussian processes, the joint distribution of $\bm Y$ and an output $f$ for an arbitrary $\bm x$ is multivariate normal,
\begin{equation}
	\begin{bmatrix}
		\bm Y^\mathsf{T} \\
		f
	\end{bmatrix}
	\sim \mathcal N \left( m \, \bm J_{j+1,1}, \,
	\begin{bmatrix}
		\bm K +\bm \Sigma & \bm k           \\
		\bm k^\mathsf{T}  & k(\bm x, \bm x)
	\end{bmatrix}
	\right),
\end{equation}
where $\bm J$ denotes an all-ones matrix, and the covariance matrices $\bm K \in \mathbb R^{j \times j}$ and $\bm k \in \mathbb R^{j \times 1}$ are
\begin{equation}
	\begin{aligned}
		\bm K_{i,i'} = k(\bm X_i, \bm X_{i'}), &  &
		\bm k_{i} = k(\bm X_i, \bm x),         &  &
		\forall i,i' \leq j.
	\end{aligned}
\end{equation}

The posterior predictive distribution for $f$ is also Gaussian,
\begin{equation}\label{fx}
	f \,|\, \bm X, \bm Y, \bm x \sim \mathcal N(\mu, \, \sigma^{2}),
\end{equation}
where the mean $\mu$ and variance $\sigma^2$ depend on previous data,
\begin{align}
	\mu      & =  \mu(\bm x) = ( \bm Y - m \, \bm J_{1,j} ) \, (\bm K+\bm \Sigma)^{-1} \, \bm k + m, \label{eq: mu_post}         \\
	\sigma^2 & =  \sigma^2(\bm x) = k(\bm x, \bm x) - \bm k^\mathsf{T} \, (\bm K+\bm \Sigma)^{-1} \, \bm k. \label{eq: sig_post}
\end{align}

\fboxsep0pt

\subsection{Data size constraining}\label{sec: limit size}
\begin{algorithm}[t]
	\caption{Adaptive search bounds}\label{alg: bounds}
	\begin{algorithmic}[1]
		\Function{Bounds}{$ \bm X, \, \bm Y, \, \bm \Sigma $}
		\State \textbf{Constant:} $ \xlbabs, \, \xubabs, \, \alpha, \, \cshrink $
		\State \textbf{Persistent:} $ \bm L, \, \bm{D_x}, \, \xbest^{k} $

		\State $\xbest^{k-1} \leftarrow \xbest^{k} $
		\State $\xbest^{k} \leftarrow \bm x_i \;\; \mathrm{s.t.} \;\; i = \argmin_i \bm \mu(\bm X_i)$ \label{eq: xbest}

		\State $ {\bm{D_x}} = \cEma \, {\bm{D_x}} + (1-\cEma) \, (\xbest^k - \xbest^{k-1}) $ \label{alg: D_x}
		\State $ \bm L \leftarrow \mathrm{sat}_{\bm{L_\mathrm{min}}}^{\bm{L_\mathrm{max}}}\Big( \cshrink \, \big(\bm L + | {\bm{D_x}} |\big) \Big) $ \label{alg: L}

		\State $ \xlb  \leftarrow \max ( \xlbabs, \ \xbest^k - \bm L ) $ \label{alg: xlb}
		\State $ \xub  \leftarrow \min ( \xubabs, \ \xbest^k + \bm L ) $ \label{alg: xub}
		\State \Return $ \xlb, \, \xub $
		\EndFunction
	\end{algorithmic}
\end{algorithm}

For the application of the optimization method for cycle-to-cycle learning type control, it is imperative to constrain the size of stored data in order to prevent the ceaseless increase of computational requirements. For that purpose, three adjustments (functions in Algorithm {\ref{alg: BO main}}) are introduced:

\begin{enumerate}[leftmargin=*]
	\item \textsc{Merge}. If several observations are performed for the same input, there is no need to store them separately. Those cost evaluations can be merged by using Bayesian inference and the equivalent cost mean and variance are obtained \cite{Moya-Lasheras2019a}.
	\item \textsc{Bounds}. The search space is made adaptive, shrinking or expanding (see Algorithm {\ref{alg: bounds}}). The space expansion in an iteration $k$ is related to the variation of $\xbest$ from the previous iteration $k-1$. First, the current one $\xbest^k$ is selected based on its mean value (see {\eqref{eq: mu_post}}). Secondly, the variations are filtered with an exponentially moving average (EMA), where $\cEma \in (0, 1)$ is the EMA coefficient. Thirdly, the bound length vector $\bm L \in \mathbb R^d$ is updated in each iteration, shrinking if ${\bm{D_x}}^k$ is small and expanding otherwise (where $\cshrink \in (0, 1)$ is the shrinkage coefficient, $| {\bm{D_x}}^k |$ is the element-wise absolute value, and $\mathrm{sat}_{\bm{L_\mathrm{max}}}^{\bm{L_\mathrm{min}}}$ denotes an element-wise saturation function between the chosen $\bm{L_\mathrm{min}}$ and $\bm{L_\mathrm{max}}$). Finally, the bounds $\xlb$ and $\xub$ are calculated around the $\xbest^k$, ensuring that they do not surpass the absolute ones $\xlbabs$ and $\xubabs$.
	\item \textsc{Remove}. Merging observations does not guarantee that the size of data history is bounded. Thus, the third measure consists in a two-step removal of points, in the case that the number of stored points $j$ surpasses a chosen limit $\jmax$. For the first step, note that the effect of points that are far away from the bounded space---depending on the characteristic length scales $\bm l$ (see \eqref{eq: kernel})---can be considered negligible. Thus, any point $\bm X_i$ that does not meet
	      \begin{equation}
		      \xlb - 3\,\bm l \leq \bm X_i \leq \xub + 3\,\bm l
	      \end{equation}
	      is removed. The shrinkage and removal of distant points are especially useful for high-dimensional inputs, in which the limited number of stored data is not enough to generate a regressor of the entire search space. However, in any case, the data size constraining is still not guaranteed. Then, as the second step, a removal criterion is defined to ensure that $j \leq \jmax$. In the previous work \cite{Moya-Lasheras2019a}, a removal criterion is proposed. The objective is to find the index $i$ which minimizes the increment of differential entropy in $\bm X_i$ due to its removal. After some algebraic simplifications, the index to be removed is obtained as
	      \begin{equation}
		      \argmin_i  \frac{\sigma^2(\bm X_i)}{\Sigma_{i,i}}
	      \end{equation}
	      where, for each point $\bm X_i$, the posterior variance $\sigma^2$ is calculated from {\eqref{eq: sig_post}}.
\end{enumerate}

\subsection{Acquisition}\label{sec: acquisition}
The final part of the algorithm is the selection of the next point $\bm x^{k+1}$ to be evaluated, which must trade off between obtaining the most information of the function (exploration) and attempting to minimize it (exploitation). As the predicted cost $f$ is a random variable, the selection of $\bm x^{k+1}$ must be carried out by the maximization of an acquisition function dependent on $\mu$ and $\sigma^2$, defined in equations \eqref{eq: mu_post} and \eqref{eq: sig_post}. The previous work \cite{Moya-Lasheras2019a} proposes a generalization of the expected improvement: the expected net improvement. First, the net improvement over the remaining $\Delta k$ commutations is defined as a piecewise function of the random variable $f \sim \mathcal N(\mu, \sigma^2)$,
\begin{equation}\label{Inet(f)}
	\srm I {net}(f) = \left\{ \begin{array}{ll}
		\Delta k \, (\fmin-f), & \mathrm{if} \ f \leq \fmin, \\
		\fmin-f,               & \mathrm{if} \ f > \fmin,
	\end{array} \right.
\end{equation}
where $\fmin = \min_i \mu(\bm X_i)$ is the best observation so far. On the one hand, an improvement over $\fmin$ means a potential improvement for the remaining $\Delta k$ commutations. On the other hand, a worsening over $\fmin$ only affects the next commutation, because in the following one it would be possible to commute with an expected cost of $\fmin$.

Then, the expected net improvement is derived, which can be expressed in relation to the expected improvement $ \E[I(f)] $ ---which is one of the most common acquisition functions---,
\begin{equation}
	\E[\srm I {net}(f)] = \fmin-\mu + (\Delta k-1) \, \E[I(f)].
\end{equation}

Note that, if $\Delta k=1$, the maximization of $\E[\srm I {net}(f)]$ is equivalent to the minimization of $\mu$. As $\Delta k$ decreases, the acquisition favors exploitation over exploration, which is the intended behavior. Also, notice that as $\Delta k$ increases, $\E[\srm I {net}(f)]/\Delta k$ tends to $\E[I(f)]$, which would correspond to a regret-free optimization. Moreover, if the number of commutations is not known, $\Delta k$ can be set as an expectation.

In this control problem, the objective is to minimize a cost related to absolute or non-negative values, e.g. contact velocities, impact sound intensities, or bouncing durations. Taking that into account, the acquisition function can be further improved. For convenience, $f$ is kept as a Gaussian random variable given by \eqref{fx} and an auxiliary variable $\fs$ is defined by saturating $f$ and ensuring its non-negativity,
\begin{equation}
	\fs = \max(f, 0).
\end{equation}
Substituting $f$ with $\fs$ in \eqref{Inet(f)}, the net improvement of the saturated $\fs$ is
\begin{equation}
	\srm I{net}(\fs) = \left\{\! \begin{array}{ll}
		\Delta k \, \fmin,     & \mathrm{if} \ f \leq 0,         \\
		\Delta k \, (\fmin-f), & \mathrm{if} \ 0 < f \leq \fmin, \\
		\fmin-f,               & \mathrm{if} \ f > \fmin,
	\end{array} \right.
\end{equation}\vspace{0mm}
and its expected value, which is the proposed acquisition function $\srm{f}{acqn}$, can be expressed in terms of the already defined $\E[\srm I{net}(f)]$,
\begin{align}
	\!\srm{f}{acqn} & = \E[\srm I{net}(\fs)]          \nonumber                                                                                                                        \\
	                & = \E[\srm I{net}(f)] - \Delta k \, \sigma \, \left( \varphi\left(-\frac{\mu}{\sigma}\right) -\frac{\mu}{\sigma} \, \Phi\left(-\frac{\mu}{\sigma}\right) \right),
\end{align}
where $\varphi$ and $\Phi$ are the standard normal probability and cumulative distribution functions, respectively. Note that the subtracting term is always positive, i.e. $\E[\srm I{net}(\fs)] < \E[\srm I{net}(f)]$, and increases with respect to the variance, prioritizing exploitation over exploration.

\section{Input parametrization}\label{sec: input}
One of the critical points of a R2R strategy is the selection of the input and its parametrization.

\subsection{System model}\label{subsec: model}
\begin{figure}
	\centering\includegraphics{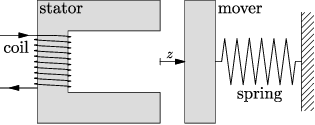}
	\caption{Schematic diagram of a reluctance actuator.}\label{fig: actuator}
\end{figure}

To justify the input selection, the dynamical system must be characterized and discussed. Thus, a generalized dynamic model for a reluctance actuator is presented. In Fig. \ref{fig: actuator}, the device is represented in its most basic form. The magnetic core is divided into a fixed part (stator) and a movable part (mover). The air gap between them is dependent on the mover position. Low-cost actuators typically have a single coil and no permanent magnets, which implies that the magnetic force has only one direction, attracting the mover toward the stator. In order to separate the mover from the stator, these devices rely on passive elastic forces. The armature position is restricted between a lower limit ($\zmin$) and an upper limit ($\zmax$).

Both the mechanical and electromagnetic dynamics must be modeled. To shorten the expressions, the time $t$ dependency and the iteration $k$ distinction are omitted. With respect to the mechanical subsystem, the motion dynamics is given by Newton's second law, with three forces,
\begin{equation}\label{eq: newton}
	\mmov \, a = \Fsp(z) + \Ff(z, v) + \Fmag(z,\phi),
\end{equation}
where $z$, $v$ and $a$ are the position, velocity and acceleration of the armature; $\Fsp$, $\Ff$, $\Fmag$ are the elastic, friction and magnetic forces; $\phi$ is the magnetic flux; and $\mmov$ is the moving mass. The only force that can be controlled---albeit indirectly---is $\Fmag$, which depends on the derivative of the gap reluctance $\Relg$ and the magnetic flux $\phi$ \cite{Ramirez-Laboreo2019b},
\begin{align}\label{eq: Fmag}
	\Fmag = - \lfrac{\dRelg(z) \, \phi^2}{2} , &  & \Relg'(z) = \lfrac{\partial \Relg(z)}{\partial z}.
\end{align}

Then, the dynamics of $\phi$ can be modeled using as input the coil current $\icoil$ or voltage $\ucoil$. The relation between the magnetic flux and the current can be expressed in terms of the reluctance, given Amp\`ere's circuital law,
\begin{equation}\label{eq: cur_flux}
	N \, \icoil + \ieddy= \big( \Relc(\phi) + \Relg(z) \big) \, \phi,
\end{equation}
where $N$ is the number of coil turns, $\Relc$ is the core reluctance, and $\ieddy$ is the net eddy current through the core. Assuming a uniform magnetic flux within the section, it can be shown {\cite{Vrijsen2014}}, {\cite{Ramirez-Laboreo2019}} that $\ieddy$ must be proportional to the time derivative of the magnetic flux,
\begin{equation}\label{eq: eddy}
	\ieddy = - \keddy \, \dot \phi.
\end{equation}

\begin{figure}[t]
	\centering
	\tikzset{elliptic state/.style={draw,ellipse,minimum height=1cm}}
	\tikzset{rectangular state/.style={draw,rectangle,minimum width=23mm,minimum height=12mm,rounded corners=4mm}}
	\begin{tikzpicture}[shorten >=1pt,node distance=4cm,on grid,auto]
		\footnotesize
		\node[rectangular state,align=center] (q_2)  {Motion \\ \vspace{-2mm} \\ $\begin{array}{l} \dot z=v \\ \dot v= f_2(z,v,\phi) \\ \dot \phi= f_3(z,\phi,u) \end{array}$};
		\node[rectangular state,align=center] (q_1) [ left=of q_2,xshift=9mm] {Lower limit \\ \vspace{-2mm} \\ $\begin{array}{l} \dot z=0 \\ \dot v= 0 \\ \dot \phi= f_3(z,\phi,u)\end{array}$};
		\node[rectangular state,align=center] (q_3) [ right=of q_2,xshift=-9mm] {Upper limit \\ \vspace{-2mm} \\ $\begin{array}{l} \dot z=0 \\ \dot v= 0 \\ \dot \phi= f_3(z,\phi,u)\end{array}$};
		\path[->]
		(q_2) edge [bend left] node [below] {$\begin{array}{r} v< 0 \; \  \wedge \; z = \zmin \\ \hspace{10mm} \Rightarrow v^+ = 0\end{array}$} (q_1)
		edge [bend right] node [below] {$\begin{array}{r} v > 0 \; \ \wedge \; z = \zmax \\ \hspace{10mm} \Rightarrow v^+ = 0\end{array}$}  (q_3)
		(q_1) edge [bend left] node [above] {$f_2(z,v,\phi)>0$} (q_2)
		(q_3) edge [bend right]  node [above] {$f_2(z,v,\phi)<0$} (q_2);
	\end{tikzpicture}
	\caption{Diagram of the hybrid automaton.} \label{fig: automaton}
\end{figure}
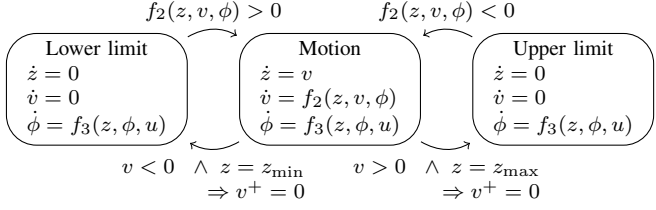

The system dynamics can be defined through a state-space representation, with three state variables: the position $z$, the velocity $v$ and the magnetic flux $\phi$. As the motion is constrained, $z$ and $v$ must be static if the armature reaches one of the two limits. Thus, the system is modeled with a hybrid automaton (Fig. {\ref{fig: automaton}}), with three discrete modes, their corresponding guard conditions, and a reset function ($v^+ =0$) in each transition from the motion mode. Most commonly, the voltage $\ucoil$ is treated as the system input $u$, because it is directly supplied to the device. In that case, the dynamics of the magnetic flux is given by the electrical circuit equation,
\begin{equation}\label{eq: electrical circuit}
	\ucoil = R \, \icoil + N \,\dot \phi.
\end{equation}

Then, substituting \eqref{eq: cur_flux} into \eqref{eq: electrical circuit} and isolating $\dot \phi$, its expression is derived,
\begin{multline}\label{eq: f_3(u)}
	\dot \phi = -\frac{R \, \big(\Relg(z)+\Relc(\phi)\big)}{N^2+R\,\keddy} \, \phi + \frac{N}{N^2+R\,\keddy} \, \ucoil.
\end{multline}

Notice the strong dependence on $R$, which can change significantly with the temperature. Thus, as a more robust alternative, this paper proposes to use the current signal as the input, $u=\icoil$, which makes the dynamics of $\phi$ invariant with respect to $R$. Then, the dynamic functions $f_2$ and $f_3$ (see Fig. \ref{fig: automaton}) are derived from \eqref{eq: newton} to \eqref{eq: eddy},
\begin{subequations}
	\begin{flalign}
		f_2(z,v,\phi)&= \frac{ \Fsp(z) + \Ff(z, v) - \dRelg(z) \, \phi^2 /2}{\mmov}, \label{eq: f2}\\[1mm]
		f_3(z,\phi,\icoil)&= -\frac{\Relc(\phi) + \Relg(z)}{\keddy} \, \phi + \frac{N}{\keddy} \, \icoil.
	\end{flalign}
\end{subequations}

\subsection{Current parametrization}\label{sec: cur_param}
The input function $\eqref{eq: U}$ must be defined, relating the decision vector $\bm x$ to the continuous input signal $u(t)$. Thus, the input must be parametrized. On the one hand, using the current as input is advantageous, as discussed in Section \ref{subsec: model}. On the other hand, its parametrization is more challenging. Regarding the voltage parametrization, bang-off-bang voltage signals can be constructed from the time intervals, which would act as the decision variables. In contrast, the current dynamics is slower, and must be taken into account. Therefore, this section presents a method of current parametrization based on the dynamic model. First, desired soft-landing position trajectories are predefined for both operations. These can be established in any way, as long as they are feasible, taking into account the physical limitations of the system. Our proposal is to use, for each operation type, the predefined position trajectory $z(t)$ as reference, modify it with computationally efficient linear transformations, and calculate the current signal $\icoil(t)$ for a given set of parameters in each iteration.

The first parameters that are considered for decision variables are the initial and final position values, $z_0$ and $\zf$. Then, the nominal position trajectory $z$ is linearly transformed into $\hat z$, so the boundary conditions $\hat z(0) = z_0$, $\hat z(\tf) = \zf$ are met,
\begin{align}
	\hat z = C_z \, \big(z - z(0)\big) + z_0, &  & C_z = \frac{\zf - z_0}{z(\tf) - z(0)},
\end{align}

The position derivatives up to the jerk $\dot a$ are necessary as well, so they must also be modified,
\begin{align}
	\hat v = C_z \, v, &  & \hat a = C_z \, a, &  & \hat{\dot a} = C_z \, \dot a.
\end{align}

Next, the current is obtained from {\eqref{eq: cur_flux}} and {\eqref{eq: eddy}}, which depend on $\phi$ and $\dot \phi$, respectively. On the one hand, the variable $\phi$ can be isolated from \eqref{eq: newton},
\begin{equation}\label{eq: phi}
	\phi = \sqrt{\frac{2\,(\Fsp(\hat z) + \Ff(\hat z, \hat v) - \mmov \, \hat a)}{\dRelg(\hat z)}}.
\end{equation}
On the other hand, $\dot \phi$ can be obtained by calculating the time derivative of \eqref{eq: newton} and isolating it, assuming that the elastic and friction functions $\Fsp$ and $\Ff$ are differentiable,
\begin{multline}\label{eq: dphi}
	\dot \phi= \frac{1}{\dRelg(\hat z)\, \phi} \left( \frac{\partial \Fsp}{\partial z}(\hat z)\,\hat v + \frac{\partial \Ff}{\partial z}(\hat z, \hat v)\,\hat v + \frac{\partial \Ff}{\partial v}(\hat z, \hat v)\,\hat a \right.\\
	\left. - \; \mmov \, \hat{\dot  a} - \frac{1}{2} \, \ddRelg(\hat z) \, \hat v \, \phi^2 \right), \hspace{6mm} \ddRelg(z) = \frac{\partial^2 \Relg(x)}{\partial z^2}.
\end{multline}
It is possible that, for a poorly chosen set of parameters and position trajectory, the radicand in {\eqref{eq: phi}} is negative in some interval. That would mean that the required magnetic force is positive (see {\ref{eq: Fmag}}), which is not physically possible. In that case, as the closest feasible approximation, $\phi$ and $\dot \phi$ are set to zero. Taken that into account, the input definition is finally expressed as
\begin{equation}\label{eq: par2u}
	\icoil =   \left\{ \begin{array}{ll} \frac{\Relc(\phi) + \Relg(\hat z)}{N}\, \phi + \frac{\keddy}{N} \, \dot{\phi}, & \mathrm{if} \ \phi^2 > 0,    \\
		0,                                                                                  & \mathrm{if} \ \phi^2 \leq 0.\end{array} \right.
\end{equation}

Thus, the current signal is a function of the position trajectory $z$, its derivatives $v$, $a$, $\dot a$, and the model parameters, including $z_0$ and $\zf$. On the one hand, the position and its derivatives are established prior to the control. Moreover, depending on the case, some of the parameters are considered constants, because they have been identified accurately and their variability between units is negligible. On the other hand, the remaining parameters are assumed unknown and will be treated as the decision variables, which can be modified in each cycle. As their magnitudes may differ greatly, it is convenient to normalize each one. Consider $\bm \theta$ to be the vector of unknown parameters. Assuming that each parameter is bounded such that $\theta_i \in [\theta^-_i, \ \theta^+_i]$, the normalization is achieved through a linear transformation,
\begin{equation}\label{eq: theta2x}
	x_i = \frac{2 \, \theta_i - \theta^+_i - \theta^-_i}{ \theta^+_i - \theta^-_i }.
\end{equation}
Thus, the decision variables are bounded $x_i \in [-1, \ 1]$.

Note that this resembles an online identification process, as the parameter-related decision variables are optimized in an iterative fashion. However, the parameters are not optimized to reduce estimation errors, but solely to improve the performance, i.e. minimize the defined cost. Thus, the parameters may not converge to their real values if it is not necessary for the control. This is not a limitation but a design choice: the objective is to minimize a certain cost, not to identify the system. Note also that, in a R2R application, the data is very limited (only one cost value per iteration), so it would not be possible to guarantee that the model parameters converge to the real values. Furthermore, the parameter-related decision variables are optimized independently for each operation, so they may be able to correct certain phenomena which are not taken into account by the dynamic model and act differently in each operation type, e.g. magnetic hysteresis.

Note that the input has only been defined for the operations, i.e. during motion. Then, the signal for the complete cycle must be constructed. After each operation, steady current values must be applied to maintain the actuator in its position ($\zmin$ after the making operation and $\zmax$ after the breaking operation). Also, there needs to be feasible transitions between the steady and the operation currents, and vice versa.

\subsection{Sensitivity analysis and dimension reduction}\label{sec: sens}
The convergence rate of the optimization process is strongly dependent on the dimension of the decision vector. Thus, it is highly recommendable to remove redundant decision variables whose effect on the input can be replicated through a combination of other variables. Some of these redundancies may be easy to spot through examination of the input function (see \eqref{eq: phi} to \eqref{eq: par2u}). In that case, through some change of variables, an equivalent input function with fewer parameters can be derived. To ensure that there are no more redundant decision variables, the input function \eqref{eq: U} can be interpreted as a discrete-time system, where the decision variables constitute the state vector,
\begin{equation}
	\left\{\begin{array}{ll}
		\bm x(t_{i+1}) & \hspace{-2mm} = \bm x(t_i),                   \\
		u(t_i)         & \hspace{-2mm} = \mathcal U( \bm x(t_i), t_i).
	\end{array}\right.
\end{equation}

Then, its local observability matrix $ \mathcal O \in \mathbb R^{d \times d}$ for consecutive time samples $t_1, \, t_2, \, \ldots, t_d$ is constructed, whose $(i,j)$th entry is $\mathcal O_{i,j} = \partial  \mathcal U(\bm x, t_i) / \partial x_j$. If the matrix is full-rank---except for singularities---, the effect of the decision variables $x_i$ are independent. Even so, it may be still possible to approximate the effect of one or several decision variables with a combination of the remaining ones. This analysis is very complex, due to the highly nonlinear input function \eqref{eq: cur_flux}. Thus, we propose to perform a local sensitivity analysis, at $\bm x=\bm 0$, which is equivalent to assuming the errors of the first-order Taylor series negligible,
\begin{align}
	\mathcal U(\bm x, t) - \mathcal U(\bm 0, t) \approx \sum_{i=1}^{d} \left.\frac{\partial \mathcal U(\bm x, t)}{\partial x_i}\right|_{\bm x = \bm 0} x_i = \nabla \mathcal U(t) \, \bm x.
\end{align}

The sensitivities are the partial derivatives of $\mathcal U$, which can be calculated for the nominal case $\bm x = \bm 0$ for every $t$. Just by comparing these partial derivatives, it is possible to determine which pair of parameters have very similar effects on the current, or which parameters have a negligible effect. Still, there may be other cases of near collinearity between sensitivities which are not apparent. One relatively simple way to analyze it is to sample the time ($t_1, \, t_2, \, \ldots$) and create a Jacobian matrix such that each $i$th row is $\nabla \mathcal U(t_i)$. Then, the rank of the matrix with a chosen tolerance can be calculated and, if the rank is not full, the decision variables to be removed can be detected with a singular value decomposition. However, this method does not take into account the bounds of each parameter. Instead, we propose the following method:

A decision variable $x_i$ is a candidate for removal if the variation of $u$ due to $x_i$ is nearly the same as a variation of $u$ with a certain $\bm x$, where its $i$th component is zero. Formally, the set of all possible solutions is
\begin{equation}\label{eq: X_minusi}
	\bm{\mathcal X}_{\setminus i} = \big\{\bm x \!\mid\! x_i = 0  \wedge  \big(\nabla \mathcal U_i(t) - \nabla \mathcal U(t) \, \bm x \big)^2 \leq {\xtol}^2 \, \forall t \big\}.
\end{equation}
where $ \xtol$ is an arbitrary tolerance. If $\bm{\mathcal X}_{\setminus i}$ is empty, the effect of the decision variable $x_i$ is not redundant and thus it should not be removed. Otherwise, $x_i$ is a candidate for removal. As the size $\bm{\mathcal X}_{\setminus i}$ may be larger than one---i.e. multiple solutions---we select the one with the smallest euclidean norm,
\begin{equation}\label{eq: x_minusi}
	\bm x_{\setminus i} = \argmin_{\bm x \, \in \, \bm{\mathcal X}_{\setminus i}} \| \bm x \|_2.
\end{equation}

This is performed for every $i$. The decision variable $x_i$ whose $\bm x_{\setminus i}$ has the minimum euclidean norm is removed. Note that, due to the removal of an input, the allowed variations of $u$ are reduced. To be able to vary $u$ as much as before the removal of $x_j$, the bounds $\bm \theta^+$ and $\bm \theta^-$ of the remaining parameters must be augmented depending on $\bm x_{\setminus j}$. Note that $\nabla \mathcal U$ is proportional to $(\bm \theta^+ - \bm \theta^-)$, so it must also be augmented. Note also that the inputs are still normalized ($x_i \in [-1, 1]$). The complete method is presented in Algorithm \ref{alg: dim red}, which must be performed for the making and breaking inputs separately.

\begin{algorithm}[t]
	\caption{Dimension reduction}\label{alg: dim red}
	\begin{algorithmic}[1]
		\Loop
		\For{$i \leftarrow 1$ \textbf{to} $d$}
		\State Calculate $\bm{\mathcal X}_{\setminus i}$ \Comment{Equation \eqref{eq: X_minusi}}
		\EndFor
		\If{$\bm{\mathcal X}_{\setminus i} = \emptyset \;\, \forall i$}
		\State \textbf{break loop}
		\EndIf
		\State Calculate $\bm x_{\setminus i}$ \Comment{Equation \eqref{eq: x_minusi}}
		\State $j \leftarrow \argmin_i \|\bm x_{\setminus i}\|_2$
		\For{$i \leftarrow 1$ \textbf{to} $d$}
		\State $ \Delta \theta \leftarrow (\theta^+_i-\theta^-_i) \, |(x_{\setminus j})_i| $
		\State $\theta^+_i \leftarrow \theta^+_i + \Delta \theta/2 $
		\State $\theta^-_i \leftarrow \theta^-_i - \Delta \theta/2 $
		\State $\nabla \mathcal U \leftarrow \nabla \mathcal U \, \big(1+|(x_{\setminus j})_i|\big)$
		\EndFor
		\State $ \left( \bm x, \, \bm{\theta^+}, \, \bm{\theta^-} \right) \leftarrow$ \Call{Remove}{$ x_j, \, \theta^+_j, \, \theta^-_j $}
		\State $d \leftarrow d -1$
		\EndLoop
	\end{algorithmic}
\end{algorithm}

\section{Analysis}\label{sec: results}
In this section, simulated and experimental results are presented and discussed.

\subsection{Solenoid valve}
\begin{figure}
	\centering
	\includegraphics[height=3cm]{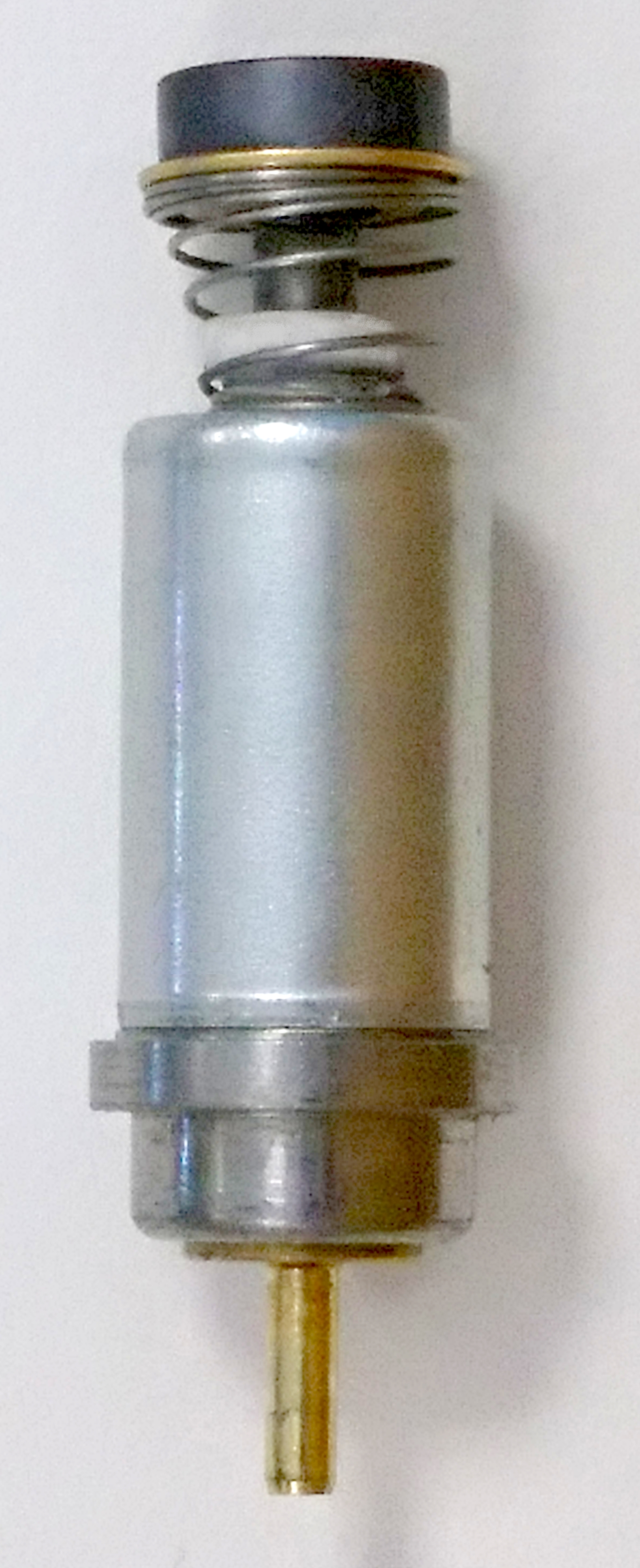} \hspace{1mm}
	\includegraphics[height=3cm]{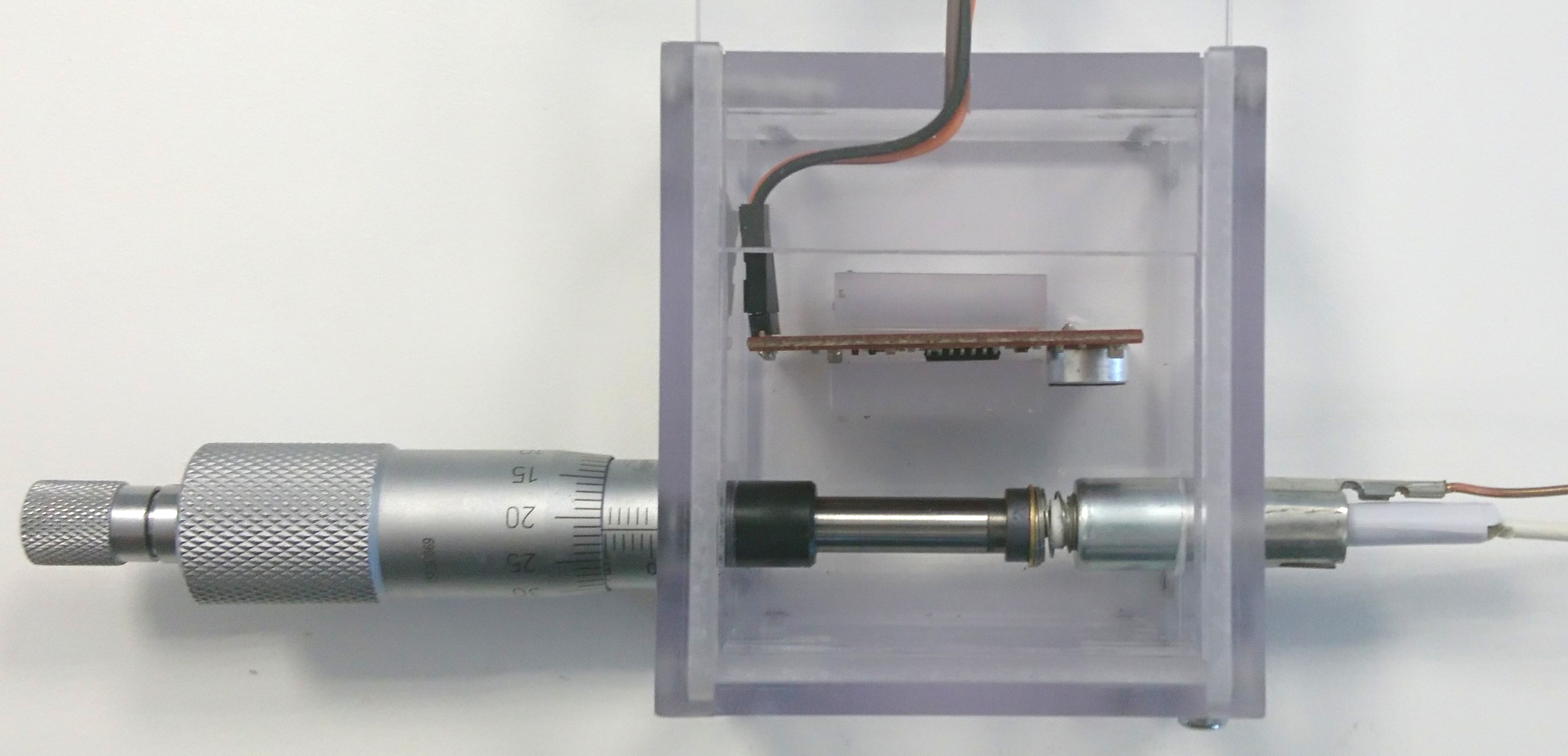}
	\caption{{Linear-travel short-stroke solenoid valve (left) and experimental setup with the valve, a micrometer to limit the maximum gap and an electret microphone to measure the impact noise (right).}}\label{fig: valve}
\end{figure}

The device used in the tests is a low-cost linear-travel solenoid valve (Fig. \ref{fig: valve}). The core is cylindrically symmetrical, with a fixed part and an armature. The coil current generates a magnetic flux through the core parts and the air gap between them. The spring force tends to open the gap, whereas the magnetic force points in the opposite direction.

Before specifying the input function, the model must be completely defined. Besides for the input parametrization, the system model is useful for performing analyses through simulations, as presented in Section \ref{subsec: sim}. The elastic force is modeled as an ideal spring, and the friction as a viscous force,
\begin{align}
	\Fsp = \ksp \, (\zsp - z), &  & \Ff = -\cf \, v,
\end{align}
being $\ksp$ the spring constant, $\zsp$ the spring resting position, and $\cf$ the friction coefficient.

In order to consider the rarely negligible magnetic saturation phenomenon, the core reluctance $\Relc$ is modeled as a parametric expression derived from the Fr\"{o}hlich-Kenelly equation,
\begin{align}
	\Relc(\phi) = \frac{k_1}{1-k_2 \, \phi}.
\end{align}

\begin{figure}[t]
	\centering
	\includegraphics{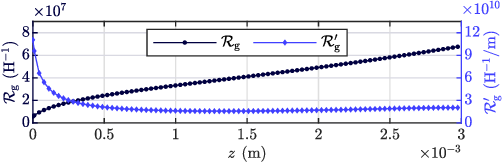}
	\caption{Gap reluctance and its derivative with respect to the gap length.}
	\label{fig: reluctance}
\end{figure}

In contrast, the gap reluctance $\Relg$ and its derivatives are obtained from lookup tables, whose data were calculated from a finite element model \cite{Ramirez-Laboreo2018}. In Fig. \ref{fig: reluctance}, $\Relg$ and $\dRelg$ are represented.

\begin{figure}[t]
	\includegraphics{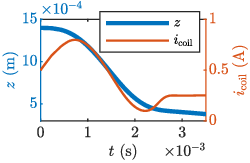} \hspace{\fill} \includegraphics{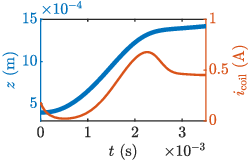}
	\caption{Desired position trajectories and nominal current signals for the making (left) and breaking (right) operations.}
	\label{fig: opt}
\end{figure}

Having defined those functions, the dynamical system is fully characterized. The model parameters are summarized in Table \ref{table: parameters}, along with their nominal values. Then, the input is parametrized following the process presented in Section \ref{sec: cur_param}. First, the desired position trajectories have been optimally calculated following the procedure presented in \cite{Moya-Lasheras2019b}, using the nominal parameter values and setting the final time to $\tf = 3.5 \; \mathrm{ms}$. Fig. \ref{fig: opt} depicts the desired position and the nominal current signals for the making and breaking operations.

\begin{table}[t]
	\centering
	\caption{Parameters of the solenoid valve} \label{table: parameters}
	
	\renewcommand{\arraystretch}{1.1} 
	\begin{tabular}{p{1.05cm} p{1.35cm} p{0.6cm}}
		\hline
		Parameter          & \multicolumn{2}{l}{Nominal value}                     \\
		\hline
		$ m_\mathrm{mov} $ & $ 1.6 \times 10^{-3}$             & $ \mathrm{kg} $   \\
		$ \ksp $           & $ 61.8$                           & $ \mathrm{N/m} $  \\
		$ \zsp $           & $ 0.0192$                         & $ \mathrm{m} $    \\
		$ \cf $            & $ 0.8$                            & $ \mathrm{Ns/m} $ \\

		$ N $              & $ 1200 $                                              \\
		\hline
	\end{tabular}
	\hspace{1mm}
	\begin{tabular}{p{1.05cm} p{1.35cm} p{0.6cm}}
		\hline
		Parameter          & \multicolumn{2}{l}{Nominal value}                        \\
		\hline
		$ \keddy $         & $ 1630$                           & $ \Omega^{-1} $      \\
		$ k_1 $            & $ 4.41 \times 10^6$               & $ \mathrm{H^{-1}} $  \\
		$ k_2 $            & $ 3.8 \times 10^4$                & $ \mathrm{Wb^{-1}} $ \\
		$ z_\mathrm{min} $ & $ 4 \times 10^{-4}$               & $ \mathrm{m} $       \\
		$ z_\mathrm{max} $ & $ 1.4 \times 10^{-3}$             & $ \mathrm{m} $       \\
		\hline
	\end{tabular}
\end{table}

Secondly, the input function is defined. To simplify the final expression, auxiliary variable $ \hat \phi = \phi / \sqrt{\mmov}$ and its derivative are used,
\begin{equation}\label{eq: par2u_v2}
	\icoil =   \left\{ \begin{array}{cc} \frac{k_1/(1- \hat k_2 \,  \hat \phi)+ \Relg(\hat z)}{ \hat N}\,  \hat \phi + \frac{\keddy}{ \hat N} \, \dot{ \hat \phi}, & {\hat \phi}^2 > 0,    \\
		0                                                                                                                              & {\hat \phi}^2 \leq 0,\end{array} \right.
\end{equation}
being $ \hat N = N/\sqrt{\mmov}$ and $ \hat k_2 = k_2 \, \sqrt{\mmov}$ auxiliary parameters. The auxiliary variable $ \hat \phi$ can be derived from \eqref{eq: phi},
\begin{equation}
	\hat \phi = \frac{\phi}{\sqrt{\mmov}} = \sqrt{\frac{2\,(\bksp\,(\zsp-\hat z) - \bcf\,\hat v - \hat a)}{\dRelg(\hat z)}},
\end{equation}
where $\bksp = \ksp/\mmov$ and $\bcf = \cf/\mmov$. Equivalently, $\dot \phi$ can be obtained from \eqref{eq: dphi},
\begin{equation}
	\dot{ \hat \phi} = \frac{\dot \phi}{\sqrt{\mmov}} = - \frac{\bksp\,\hat v + \bcf\,\hat a + \hat{\dot  a} + \ddRelg(\hat z)\,\hat v\,{\hat\phi}^2/2}{\dRelg(\hat z)\, \hat\phi}.
\end{equation}
Thus, the input depends on these parameters:
\begin{equation}\label{eq: theta}
	\bm \theta = \begin{bmatrix}
		z_0 & \zf & \bksp & \zsp & \bcf & \hat N & k_1 & \hat k_2 & \keddy
	\end{bmatrix}^\mathsf T,
\end{equation}
which then are normalized (see \eqref{eq: theta2x}) to obtain the decision vector $\bm x$. The parameter bounds are set to $\pm 10\%$ of the nominal values $\thetanomi$. Specifically,
\begin{align}
	\theta_i^\pm = \thetanomi \pm 0.1 \, {\Delta z}^\mathrm{nom}, &  & {\Delta z}^\mathrm{nom} =  {\zmax}^{\hspace{-3mm}\mathrm{nom}} -  {\zmin}^{\hspace{-3mm}\mathrm{nom}},
\end{align}
if $\theta_i$ is $z_0$ or $\zf$. Otherwise,
\begin{align}
	\theta_i^\pm = \thetanomi \pm 0.1 \, \thetanomi.
\end{align}

Note there was a redundant degree of freedom related to $\mmov$, and has been removed thanks to the change of variables \eqref{eq: par2u_v2} and the definition of auxiliary parameters \eqref{eq: theta}. Theoretically, there are no other redundancies, because the observability matrix $\mathcal O$---as defined in Section {\ref{sec: sens}}---is full-rank. However, there may be some decision variables whose effect can be approximated from a combination of the others. Thus, in order to detect other candidates for removal, we use Algorithm \ref{alg: dim red}, which was presented in Section \ref{sec: sens}. The tolerance is set to $ \xtol = 10^{-3} \; \mathrm{A}$. Then, the decision variables to be removed are the ones related to $\bksp$, $k_1$ and $\keddy$, for both operation types. Therefore, those parameters are kept constants, and the bounds of the rest of parameters are augmented to compensate for the removal of degrees of freedom. The dimension of the new decision vectors is $d=6$.

\subsection{Compared strategies}
The proposed R2R control based on Bayesian optimization (R2R-BO) is compared with two alternatives. They use the same input definition, but different optimization methods. The first R2R strategy is based on the pattern search method (R2R-PS), previously proposed for soft landing of actuators \cite{Ramirez-Laboreo2017a}, and requires evaluating $2 \, d + 1$ points in each iteration (one being the previous best point and the other being around it). The shrink and expand coefficients are set to $1/2$ and $2$ respectively, i.e. the mesh size is halved or doubled if the new best point is the same or not, respectively, as the previous one. The starting meshes are set equal for both operations,
\begin{equation}
	\bm X_{2\,d+1} = \begin{bmatrix} \bm 0 & \bm I & -\bm I \end{bmatrix}.
\end{equation}

The second strategy is based on the Nelder-Mead method (R2R-NM), which is also a direct search method \cite{Nelder1965}. Compared to the pattern search, it requires in practice fewer evaluations per iteration. It has been previously applied for other types of learning controllers, e.g. a cyclic adaptive feedforward approach controller for solenoid valves in internal combustion engines \cite{Tsai2012}. For its application in R2R control, several modifications are introduced to the algorithm: the definition of a minimum simplex volume $\Vmin$, the modification of the shrink step, and the rearrangement of the algorithm to allow only one evaluation per iteration. For more information about the algorithm, see Appendix \ref{sec: nelder-mead}. The initial points are set such that they form a regular simplex, centered in $\bm 0$, with every vertex at a unit distance from it, and randomly rotated. The method also depends on three constant coefficients for the reflection, expansion and contraction steps ($\creflect$, $\cexpand$ and $\cexpand$ respectively). The constants of the method are set as follows:
\begin{align}
	\Vmin = 0.005^d, &  & \creflect = 1, &  & \cexpand = 2, &  & \ccontract = 0.5.
\end{align}

With respect to our main proposal R2R-BO, a set of $2\,d+1$ fixed initial points ($\bm X_{2\,d+1} = \begin{bmatrix} \bm 0 & \bm I & -\bm I \end{bmatrix}$) are evaluated, gaining some knowledge across the search space before starting the optimization. Regarding the Gaussian process regression, the prior mean values and kernel hyperparameters from \eqref{eq: kernel} are specified for each case so the optimization process works efficiently. Then, the following variables are initialized:
\begin{align}
	\bm L^0 = \bm J_{d,1}, &  & {\bm{D_x}}^0 = \bm 0.
\end{align}

Several combinations of the constants $\cshrink$ and $\cEma$ have been tested in a preliminary simulated analysis. The selected ones for the final comparison are:
\begin{align}
	\cshrink = 0.98, &  & \cEma = 0.9.
\end{align}

The remaining constants are set:
\begin{equation}
	\begin{aligned}
		\jmax & = 50, & \bm{L_{\mathrm{min}}} & = 0.001 \, \bm J_{d,1}, & \bm{L_{\mathrm{max}}} & = 2 \, \bm J_{d,1}.
	\end{aligned}
\end{equation}
The number of stored data $\jmax$ depends on the implementation, but should be chosen as large as possible. The maximum bound length $\bm{L_{\mathrm{max}}}$ should be selected conservatively, ensuring that any point is reachable from any other. On the other hand, $\bm{L_{\mathrm{min}}}$ should be small enough to ensure that the smallest region can be properly approximated with $\jmax$ points.

\subsection{Simulations}\label{subsec: sim}
The strategies are compared through a Monte Carlo method, performing 500 simulations of 200 making and breaking commutations for each case. The simulations depend on the parameter vector $\bm p$, which is defined as
\begin{equation}
	\hspace{-1.5mm} \bm p = \big[
		\zmin \hspace{1.8mm}  \zmax \hspace{1.8mm}  \ksp \hspace{1.8mm}  \zsp \hspace{1.8mm}  \cf \hspace{1.8mm}  N \hspace{1.8mm}  k_1 \hspace{1.8mm} k_2 \hspace{1.8mm}  \keddy \hspace{1.8mm}  \mmov \big]^\mathsf T.
\end{equation}

To emulate variability, each model parameter $p_i$ is perturbed in every commutation. Thus, for each $k$th commutation of the $n$th simulation, the parameter $p_i^{n,k}$ is defined as a random deviate. Specifically,
\begin{align}
	p_i^{n,k} \sim \mathcal{N}( \bar{p}_i^{\hspace{0.3mm} n}, \,  ({\Delta z}^\mathrm{nom} \, \sigtheta)^2)
\end{align}
if $p_i$ is $\zmax$ or $\zmin$. Otherwise,
\begin{equation}
	p_i^{n,k} \sim \mathcal{N}( \bar{p}_i^{\hspace{0.3mm} n}, (\pnomi \, \sigtheta)^2),
\end{equation}
where the $\pnomi$ is the corresponding nominal parameter (Table \ref{table: parameters}). The relative standard deviation $\sigtheta$ serves to emulate the cycle-to-cycle perturbation. Additionally, to emulate unit-to-unit variability, the mean $\bm{\bar{p}}^{\hspace{0.3mm} n}$ is randomly modified in each simulation $n$, given a continuous uniform distribution,
\begin{equation}
	\bar{p}_i^{\hspace{0.3mm} n} \sim \mathrm{unif}(\pnomi - 0.07 \, {\Delta z}^\mathrm{nom}, \, \pnomi + 0.07 \, {\Delta z}^\mathrm{nom} ),
\end{equation}
if $p_i$ is $\zmax$ or $\zmin$. Otherwise,
\begin{equation}
	\bar{p}_i^{\hspace{0.3mm} n} \sim \mathrm{unif}(0.93 \, \pnomi, \, 1.07 \, \pnomi).
\end{equation}

\begin{figure}[t]
	\vspace{-1mm}
	\centering
	\subfloat[Making ($\sigtheta = 10^{-3}$).	\label{fig: make1}]	{\includegraphics{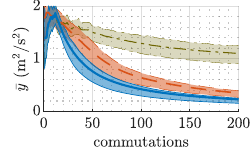}} \hspace{\fill}
	\subfloat[Breaking ($\sigtheta = 10^{-3}$).	\label{fig: break1}]	{\includegraphics{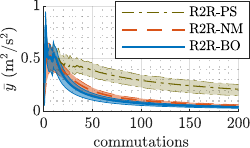}} \\ \vspace{-0.0mm}
	\subfloat[Making ($\sigtheta = 2\times10^{-3}$).	\label{fig: make2}]	{\includegraphics{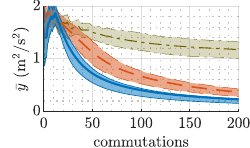}} \hspace{\fill}
	\subfloat[Breaking ($\sigtheta = 2\times10^{-3}$).	\label{fig: break2}]	{\includegraphics{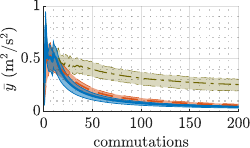}} \\ \vspace{-0.0mm}
	\subfloat[Making ($\sigtheta = 5\times10^{-3}$).	\label{fig: make3}]	{\includegraphics{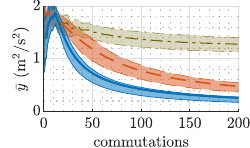}} \hspace{\fill}
	\subfloat[Breaking ($\sigtheta = 5\times10^{-3}$).	\label{fig: break3}]	{\includegraphics{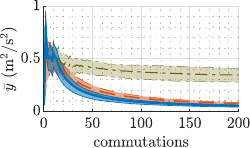}} \\ \vspace{-0.0mm}
	\subfloat[Making ($\sigtheta = 10^{-2}$).	\label{fig: make4}]	{\includegraphics{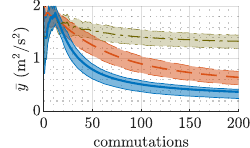}} \hspace{\fill}
	\subfloat[Breaking ($\sigtheta = 10^{-2}$).	\label{fig: break4}]	{\includegraphics{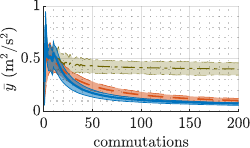}}
	\caption{Comparison of results (mean values and $25$th-$75$th percentile intervals) from R2R-PS, R2R-NM and R2R-BO, for different $\sigtheta$.}	\label{fig: compare_sim}
\end{figure}

The objective is to minimize the impacts. Therefore, the output or cost $y$ is defined as the sum of squared velocities during contact. Note that the elastic bouncing phenomenon is not considered in the proposed model, but bouncing will still occur if the impact acceleration has opposite sign to the impact velocity. Then, for every simulation, the average cost $\bar y$ is calculated for each number of commutations $k$, $\bar y(k) = \sum_{i=1}^k y^i / k$. This way, we determine how good each control is for any number of commutations, and how it improves as the number of commutations increases. In Fig. \ref{fig: compare_sim}, the mean and $25$th-$75$th percentile intervals of $\bar y$ are represented as functions of the number of commutations. Figs. \ref{fig: make1} and \ref{fig: break1} show the average costs for $\sigtheta = 10^{-3}$. Then, $\sigtheta$ is increased to $2\times10^{-3}$ (Figs. \ref{fig: make2} and \ref{fig: break2}), $5 \times 10^{-3}$ (Figs. \ref{fig: make3} and \ref{fig: break3}), and $10^{-2}$ (Figs. \ref{fig: make4} and \ref{fig: break4}). R2R-BO and R2R-PS has the same 13 starting points, so their costs are the same up until that point (the R2R-NM results at the start are also very similar). However, from that point forward the costs from the different strategies start to diverge: the R2R-BO costs are the smallest, following by the costs from R2R-NM, which is still much better than the R2R-PS. As expected, all costs get worse as $\sigtheta$ increases, and the difference between R2R-BO and R2R-NM augments. In the breaking operations, the costs are generally smaller than in the making operations, and the improvement of R2R-BO over R2R-NM is less significant.

On the one hand, note that R2R-PS and R2R-NM do not require any knowledge of the black-box functions, which makes their implementation for different actuators more straightforward than R2R-BO. They are also more computationally efficient. Between those two, the best strategy is clearly R2R-NM. On the other hand, the best results are obtained with R2R-BO. Moreover, R2R-BO takes into account the stochasticity, so the improvement over R2R-NM is more appreciable for larger cycle-to-cycle variabilities.

\subsection{Experimental results}
To validate the R2R control, experimental testings are performed with five solenoid valves. They are supposedly equal to the one used for identification (see Section \ref{subsec: model}), but unit-to-unit variability is expected. Each control strategy is tested ten times for each device. The impact velocities are unavailable, due to the lack of a position sensor. Instead, an electret microphone is used to measure the impact sound (Fig.~\ref{fig: valve}). The cost is then calculated from the microphone voltage signals,
\begin{equation}
	y = \int_{t_0}^{t_0 + \Delta t} {u_\mathrm{audio}}^2(t) \, \mathrm dt,
\end{equation}
where $t_0$ is established as the first instant $t$ where $ u_\mathrm{audio}(t) > \mathrm{max}( u_\mathrm{audio})/5$ and $\Delta t = 0.01 \ \mathrm s$.

\begin{figure}
	\centering
	\subfloat	{\includegraphics{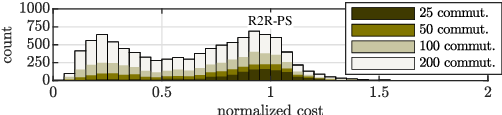}} \\ \vspace{-2.75mm}
	\subfloat	{\includegraphics{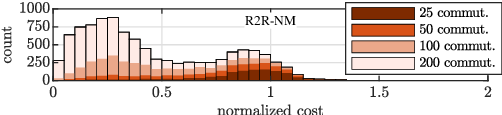}} \\ \vspace{-2.75mm}
	\subfloat	{\includegraphics{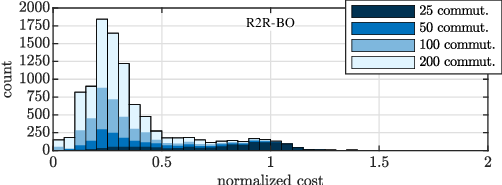}}
	\caption{Histograms of normalized costs for the first 25, 50, 100 and 200 commutations of each control.}	\label{fig: compare_exp}
\end{figure}

For reference, a non-controlled case is also evaluated, with squared voltage signals of $0$ and $60 \ \mathrm V$. Then, the controlled costs $y$ are divided by the average non-controlled one, resulting in normalized costs. This analysis is focused on the making operations, which present the most notable impact noises. The normalized costs of the making operations are summarized in the histograms from Fig. \ref{fig: compare_exp}.

If the control is applied for only 25 commutations, there is a slight improvement over the non-controlled case: in average $13\%$ (R2R-PS), $15\%$ (R2R-NM) and $25\%$ (R2R-BO). As expected, the results improve as the number of commutations increases. For 200 commutations, the average improvements are $35\%$ (R2R-PS), $44\%$ (R2R-NM) and $64\%$ (R2R-BO). Note that the results of R2R-NM are worse than expected from the simulated analysis. In the fist 100 commutations, the difference between R2R-PS and R2R-NM is not very significant. This indicates that the strategy has a slower convergence rate when applied to the real system. Nevertheless, equivalently to the simulated results, the worst strategy is R2R-PS. Note also that R2R-NM is able to obtain normalized costs below $0.10$ more often than R2R-BO, but it also obtains normalized costs larger than $0.35$ much more frequently. R2R-BO, on the other hand, is more conservative, keeping most normalized costs around $0.25$. Thus, its results are the most robust. Moreover, in average, R2R-BO is better than the other two alternatives for any number of commutations.

\section{Conclusions}\label{conclusions}
This paper presents a new run-to-run strategy based on Bayesian optimization for soft-landing control of short-stroke reluctance actuators. The control does not use position feedback, which makes it useful for applications in which position sensors or observers are not feasible. The complete algorithm has been separated into different parts, most notably the input definition and the search algorithm.

The current has been selected as the input and an input function has been defined. The input function is model-based, so it requires prior knowledge of the system dynamics. Methods for current parametrization and parameter reduction have been proposed for simple reluctance actuators, and they have been put into effect with the dynamic model of a plunger-type solenoid valve. Using the current as input makes the position dynamics independent on the coil resistance. Thus, the controller is more robust to temperature changes. This is an improvement over previous works \cite{Ramirez-Laboreo2017a, Moya-Lasheras2019a}, which propose to use the voltage as input.

Although the input definition is based on a model, the cost function is still a black box. Thus, the cycle-to-cycle search process is a black-box optimization. The proposed algorithm is completely generalized. It is based on Bayesian optimization, which has been adapted in several ways for its application in the run-to-run control. The proposed control (R2R-BO) has been compared with two alternatives, that uses direct search methods (R2R-PS and R2R-NM). One important advantage of R2R-BO is that, as it uses Gaussian process regressors, it directly accounts for uncertainty and hence it is more robust. Besides, it efficiently exploits previous data to select new points and converge rapidly to an optimal solution. The simulated and experimental results show the improvement of R2R-BO with respect to the other alternatives.

\appendices
\section{Alternative search method}\label{sec: nelder-mead}
For comparison purposes, an alternative search method has been developed. It is based on the Nelder-Mead simplex method, which is one of the most widely used direct search methods. The algorithm keeps a set of $ d + 1 $ evaluated points (where $ d $ is the dimension of the search space) forming the vertices of a non-degenerate simplex. In each iteration, one or more tasks are performed, modifying the vertices and reevaluating the function in those new points. The tasks are:
\begin{enumerate}[leftmargin=*]
	\item The worst vertex (the point which corresponds to the largest function evaluation) is reflected with respect to the centroid of the remaining vertices.
	\item If the reflected point is the best so far, the reflected point is expanded farther from the centroid. The best one of these two is kept.
	\item If the reflected vertex is the worst point, a contraction is performed toward the centroid.
	\item If the contracted point is worse than its original, all vertices of the simplex are modified and evaluated.
\end{enumerate}

In order to implement it in the R2R control, the standard algorithm is modified in several ways. The first modification is related to the last task. The standard method shrinks the simplex, fixed on the best vertex. As stated by various authors \cite{Ernst68,Nelson1991}, the shrinkage is potentially problematic because it may rapidly converge to a non-optimum. It is also inadvisable for optimizing stochastic functions, as the best point is never reevaluated. \cite{Ernst68} proposed a translation such that the previous best vertex becomes the center of the new one. In our proposal, the simplex is also centered at the best point. However, instead of maintaining its shape, the vertices are recalculated so they form a randomly rotated regular simplex with the same volume. This prevents both simplex degeneracy and retention of misleadingly good evaluations due to noisy measurements.

Secondly, instead of setting a termination condition, the algorithm is designed to continually operate. Specifically, instead of terminating the optimization when the simplex volume is lesser than a chosen minimum $\Vmin$, the simplex is prevented to be contracted below that threshold. The simplex volume $V$ is updated in each iteration, augmenting it when expanding and reducing it when contracting, without requiring to compute it from the vertices in every iteration. Also, the actual simplex volume is not important, it is possible to initialize it to $1$ and then set $\Vmin$ in accordance.

The function is described in Algorithm \ref{alg: nelder}. Its inputs are the point $\bm x^k$, which was obtained in the previous iteration, and its evaluation $y^k$. Its output is the next point $\bm x^{k+1}$. In the process, several variables are updated inside the function: the simplex vertices $\bm X \in \mathbb R^{d \times (d+1)}$ and their respective evaluations $\bm Y \in \mathbb R^{1 \times (d+1)}$, the centroid $\xc \in \mathbb R^d$, and the volume $V$. Normally, the Nelder-Mead method is presented in a sequential way with multiple evaluations per iteration. However, as stated in Algorithm \ref{alg: main}, only one evaluation is needed in each function call. Thus, the current step (reflecting, expanding, contracting or rearranging) is stored in the persistent variable $s \in \left\{1, 2, 3, 4\right\}$. For the same reason, the index of the vertex to be evaluated is stored in the persistent variable $j \in \left\{1, 2, \ldots, (d+1)\right\}$. Note that all these variables are different for each operation type but, for the sake of simplicity, that distinction is omitted. The method also depends on three constant coefficients for the reflection, expansion and contraction steps ($\creflect$, $\cexpand$ and $\cexpand$ respectively).

\begin{algorithm}
	\caption{Nelder-Mead optimization}\label{alg: nelder}
	\begin{algorithmic}[1]
		\Function{Search}{$\bm x^k, y^k$}
		\State \textbf{Constant:} $ d, \, \creflect, \, \cexpand, \, \ccontract, \, \Vmin $
		\State \textbf{Persistent:} $ \bm X, \, \bm Y, \, \xc, \, V, \, s, \, j $
		\Switch{$s$} \Comment{Update simplex $\bm X, \bm Y$ and step $s$}
		\Case{$1$} \LeftComment{Reflect}
		\If{$ y^k \leq Y_{d+1}$}
		\State $ \left(\bm X_{d+1}, \, Y_{d+1}\right) \leftarrow \left( \bm x^k, \, y^k \right) $
		\EndIf
		\If{$ y^k < Y_1 $}
		\State $ s \leftarrow 2 $
		\ElsIf{$ y^k > Y_d $ \textbf{and} $ \ccontract \, V \geq \Vmin $}
		\State $ s \leftarrow 3 $
		\ElsIf{$ y^k > Y_d $}
		\State $ \bm X \leftarrow $\Call{Create Simplex}{$ \bm X_1, V$}
		\State $ j \leftarrow 1 $ \Comment{Initialize index of simplex vertex}
		\State $ s \leftarrow 4 $
		\EndIf
		\EndCase
		\Case{$2$} \LeftComment{Expand}
		\If{ $ y^k < Y_{d+1} $}
		\State $ \left(\bm X_{d+1}, \, Y_{d+1}\right) \leftarrow \left( \bm x^k, \, y^k \right) $
		\EndIf
		\State $s \leftarrow 1$
		\EndCase
		\Case{$3$} \LeftComment{Contract}
		\If{ $ y^k < Y_{d+1} $}
		\State $ \left(\bm X_{d+1}, \, Y_{d+1}\right) \leftarrow \left( \bm x^k, \, y^k \right) $
		\State $s \leftarrow 1$
		\Else
		\State $ \bm X \leftarrow $\Call{Create Simplex}{$ \bm X_1, V$}
		\State $j \leftarrow 1$ \Comment{Initialize index of simplex vertex}
		\State $s \leftarrow 4$
		\EndIf
		\EndCase
		\Case{$4$} \LeftComment{Simplex}
		\State $ Y_j \leftarrow  y^k $
		\If{$ j = d+1$}
		\State $ s \leftarrow 1 $
		\EndIf
		\EndCase
		\EndSwitch

		\Switch{$s$} \Comment{Find next decision vector $ \bm x^{k+1} $}
		\Case{$1$} \LeftComment{Reflect}
		\State $ \left( \bm X, \bm Y \right) \leftarrow $\Call{Sort}{$ \bm Y $} \Comment $Y_1<Y_2<\ldots<Y_{d+1}$
		\State $ \xc \leftarrow \sum_{j=1}^d \bm X_{j} / d $ \Comment{Centroid of $\bm X_1, \ldots ,\bm X_d$}
		\State $ \bm x^{k+1} \leftarrow \xc + \creflect \left( \xc - \bm X_{d+1} \right) $
		\EndCase
		\Case{$2$} \LeftComment{Expand}
		\State $ \bm x^{k+1} \leftarrow \xc + \cexpand \, \left ( \bm X_{d+1} - \xc \right)$
		\State $ V \leftarrow \cexpand \, V $ \Comment{ Update volume}
		\EndCase
		\Case{$3$} \LeftComment{Contract}
		\State $ \bm x^{k+1} \leftarrow \xc + \ccontract \, \left ( \bm X_{d+1} - \xc \right)$
		\State $ V \leftarrow \ccontract \, V $ \Comment{ Update volume}
		\EndCase
		\Case{$4$} \LeftComment{Simplex}
		\State $ \bm x^{k+1} \leftarrow \bm X_{j} $
		\State $j \leftarrow j+1$ \Comment{Next simplex vertex}
		\EndCase
		\EndSwitch

		\State \Return $ \bm x^{k+1} $
		\EndFunction
	\end{algorithmic}
\end{algorithm}

\bibliographystyle{IEEEtran}

\end{document}